\documentclass[a4paper,12pt]{article}
\usepackage{amssymb,amsmath,amsthm,graphics}
\usepackage{hyperref} 

\newcommand{\doiref}[2]{\href{http://dx.doi.org/#1}{#2}}
\newcommand{\arxivref}[2]{\href{http://arxiv.org/abs/#1}{arXiv:#1.}}

\newcommand{\RR}{{\mathbb{R}}}

\newcommand{\NN}{{\mathbb{N}}}
\newcommand{\pa}{\partial}
\newcommand{\ii}{{\rm i}}
\newcommand{\dd}{{\rm d}}

\newcommand{\Tr}{\mathrm{tr}}

\theoremstyle{plain}

\theoremstyle{definition}

\title{Topological energy bounds for\\frustrated magnets}
\author{Derek Harland\footnote{email address: d.g.harland@leeds.ac.uk}
  \bigskip
  \\School of Mathematics,
  \\University of Leeds,
  \\LS2 9JT}
\date{16th March 2019}

\begin{document}

\maketitle

\abstract{Frustrated magnets are known to support two-dimensional topological solitons, called skyrmions.  A continuum model for frustrated magnets has recently been shown to support both two-dimensional skyrmions and three-dimensional knotted solitons (hopfions).  In this note we derive lower bounds for the energies of these solitons expressed in terms of their topological invariants.  The bounds are linear in the degree in the case of skyrmions and scale as the Hopf degree to the power 3/4 in the case of hopfions.}

\section{Introduction}

Since their experimental observation \cite{yokphmnt10} in 2009, there has been intense interest in magnetic skyrmions.  These skyrmions were predicted to exist in \cite{by89} and owe their stability to the Dzyaloshinskii-Moriya (DM) interaction.  An alternative stabilising mechanism, realised in recent experiments \cite{sub10nm}, is provided in frustrated magnets.  Theoretical studies of frustrated magnets have revealed a rich variety of phenomena \cite{ock12,lm15,lh16,sutcliffe17}.  In \cite{lh16} a Ginzburg--Landau type continuum model of frustrated magnets was developed, and this continuum model has been shown to support not only two-dimensional skyrmions \cite{lh16} but also three-dimensional knotted solitons \cite{sutcliffe17}.


Topological energy bounds are a ubiquitous feature in the theory of topological solitons \cite{mantonsutcliffe}.  They give an indication as to how the energy of a soliton depends on its topological complexity, and can sometimes give information about the stability of solitons.  They can provide useful consistency checks on numerical methods.  They can also be useful in mathematical proofs of existence of solitons.

In models that support two-dimensional skyrmions, one often finds linear energy bounds of the form $E\geq CN$, where $E$ and $N$ are the energy and degree of a skyrmion, and $C$ is a positive constant.  Bounds of this type have been derived for the O(3) sigma model \cite{bp75}, the baby Skyrme model \cite{pz95}, and chiral magnetic skyrmions stabilised by the DM interaction \cite{melcher14} (see also \cite{brs18} for recent developments).  Linear bounds support a particle-like interpretation of skyrmions, with the mass (energy) roughly proportional to the number of particles (degree).  They also give information about stability to separation.  If the energy $E_1$ of a one-soliton is close to its bound, $E_1-C\ll C$, then the energy $NE_1-E_N$ required to separate an $N$-soliton of energy $E_N$ into $N$ one-solitons is less than or equal to $N(E_1-C)\ll NC$.  Thus in systems where the bound is almost saturated, it is relatively easy to separate an $N$-soliton into $N$ one-solitons.

Three-dimensional solitons in magnetic systems are characterised by the Hopf degree $Q\in\pi_3(S^2)\cong\mathbb{Z}$.  In such systems one often finds energy bounds of the form $E\geq CQ^{3/4}$.  The first bound of this type was obtained for the Faddeev model \cite{vk79}, but more recently bounds have been obtained for several variant models, including the Aratyn-Ferreira-Zimerman model, the Nicole model, and a Faddeev-type model stabilised by a potential term \cite{harland14}.  The sub-linear power $\frac34$ is characteristic of knotted solitons, and reminiscent of the lower bound on the ropelength energy of a knot \cite{bucksimon99}, which scales as crossing number to the power $\frac34$.  It also suggests a high degree of stability: if the energy of a $Q$-soliton grows as $Q^{3/4}$, then it is energetically very expensive for a $Q$-soliton to separate into solitons of lower charge.

In this note we will derive topological energy bounds for two- and three-dimensional solitons in the Ginzburg--Landau model of frustrated magnets developed by Lin and Hayami \cite{lh16}.  These scale as $N$ and $Q^{3/4}$ respectively.  We test the sharpness of the two-dimensional bound by computing energies of one-solitons and comparing them with the lower bound, and show that the $\frac34$ power in the three-dimensional bound is optimal.

\section{The energy and its lower bound}

Our starting point is the energy functional derived by Lin and Hayami \cite{lh16}:
\begin{equation}
E[\mathbf{m}] = \int \bigg[ -\frac{I_1}{2}|\nabla\mathbf{m}|^2 + \frac{I_2}{2}|\triangle\mathbf{m}|^2 + (H-\mathbf{H}.\mathbf{m}) \bigg] \dd^n\mathbf{x}.
\end{equation}
Here $\mathbf{m}(\mathbf{x})$ is a three-component vector-valued function satisfying the constraint $\mathbf{m}.\mathbf{m}=1$.  The vector $\mathbf{H}$ describes the external magnetic field.  We assume throughout that $H=|\mathbf{H}|\geq\frac{1}{4}$; in this situation the vacuum for the theory is $\mathbf{m}=\mathbf{H}/H$ \cite{lh16}.  The constant $H$ has been included in the energy density so that the vacuum has energy zero; effectively, we have subtracted the vacuum energy from the total energy.  As in \cite{lh16,sutcliffe17} we choose units so that $I_1=I_2=1$.  We also apply a rotation so that $\mathbf{H}=H\mathbf{e}_3=(0,0,H)$.  With these choices the energy becomes
\begin{equation}
\label{energy}
E[\mathbf{m}] = \int \bigg[ -\frac{1}{2}|\nabla\mathbf{m}|^2 + \frac{1}{2}|\triangle\mathbf{m}|^2 + H(1-m_3) \bigg] \dd^n\mathbf{x}.
\end{equation}
We will be interested in spatially localised solitons for this energy.  We therefore assume that the field $\mathbf{m}$ far from the soliton lies in the vacuum, that is,
\begin{equation}
\label{bc}
\mathbf{m}(\mathbf{x})\to\mathbf{e}_3\text{ and }|\nabla\mathbf{m}(\mathbf{x})|\to0\text{ as }|\mathbf{x}|\to\infty.
\end{equation}

The key identity that we use to obtain lower bounds on the energy is the following, valid for any $t_1,t_2\in\RR$:
\begin{multline}
\int \frac{1}{2}|t_1\triangle\mathbf{m}+t_2(\mathbf{m}-\mathbf{e}_3)|^2\dd^n\mathbf{x} \\
= \int \bigg[ \frac{t_1^2}{2}|\triangle\mathbf{m}|^2 - t_1t_2|\nabla\mathbf{m}|^2 + t_2^2(1-m_3)\bigg]\dd^n\mathbf{x}.
\end{multline}
Here we have used an integration by parts.  The associated boundary term vanishes due to our assumptions \eqref{bc}.  We have also simplifed the non-derivative terms using the equations $|\mathbf{m}|^2=|\mathbf{e}_3|^2 = 1$.

Subtracting this positive quantity from the energy yields the inequality
\begin{align}
E &\geq E - \int \frac{1}{2}|t_1\triangle\mathbf{m}+t_2(\mathbf{m}+\mathbf{e}_3)|^2\dd^n\mathbf{x} \\
&= \int \bigg[ \frac{1}{2}(1-t_1^2)|\triangle\mathbf{m}|^2 + \left(t_1t_2-\frac{1}{2}\right)|\nabla\mathbf{m}|^2 \nonumber \\
& \qquad\qquad\qquad + (H-t_2^2)(1-m_3)\bigg]\dd^n\mathbf{x}.
\label{crucial step}
\end{align}
The lower bound is non-negative as long as $t_1,t_2$ are chosen so that
\begin{equation}
\label{constraints}
0\leq t_1\leq 1,\quad t_1t_2\geq\frac{1}{2},\quad 0\leq t_2\leq\sqrt{H}.
\end{equation}
These constraints have solutions as long as $H\geq\frac14$.  From now on we assume that $t_1,t_2$ satisfy inequalities \eqref{constraints}.

A slightly more useful lower bound can be obtained by rewriting the term $|\triangle\mathbf{m}|^2$.  Since $\mathbf{m}$ is a unit vector,
\begin{equation}
|\triangle\mathbf{m}|^2 \geq (\mathbf{m}.\triangle\mathbf{m})^2 = |\nabla\mathbf{m}|^4.
\end{equation}
Here we have used the identity $0=\frac{1}{2}\triangle|\mathbf{m}|^2 = \mathbf{m}.\triangle\mathbf{m}+|\nabla\mathbf{m}|^2$.

To further simplify this term, consider the $3\times3$ matrix $M_{ij}=\pa_km_i\pa_km_j$.  This matrix is symmetric, and its eigenvalues are non-negative because it can be written as the product of a matrix with its transpose.  One of its eigenvalues is zero, because $M_{ij}m_j = 0$ due to the identity $0=\frac12\pa_k|\mathbf{m}|^2=\mathbf{m}.\pa_k\mathbf{m}$.  We denote the remaining eigenvalues by $\lambda_1^2,\lambda_2^2$.  Then
\begin{multline}
\label{E4 bound}
|\nabla\mathbf{m}|^4 = (\Tr M)^2 = (\lambda_1^2+\lambda_2^2)^2 \geq 4\lambda_1^2\lambda_2^2 = 2((\Tr M)^2-\Tr(M^2)) \\
= 2\sum_{i,j=1}^n\Big[|\pa_i\mathbf{m}|^2|\pa_j\mathbf{m}|^2-(\pa_i\mathbf{m}.\pa_j\mathbf{m})^2\Big] \\
= 2\sum_{i,j=1}^n|\pa_i\mathbf{m}\times\pa_j\mathbf{m}|^2 = 2\sum_{i,j=1}^n(\mathbf{m}.\pa_i\mathbf{m}\times\pa_j\mathbf{m})^2.
\end{multline}
The last of these identities follows from the fact that $\mathbf{m}$ is perpendicular to the derivatives of $\mathbf{m}$.  Using this, we arrive at
\begin{align}
\label{general bound}
E&\geq \int \bigg[ 2(1-t_1^2)\mathcal{E}_4 + \left(t_1t_2-\frac{1}{2}\right)\mathcal{E}_2 + (H-t_2^2)\mathcal{E}_0\bigg]\dd^n\mathbf{x}, \\
\mathcal{E}_4 &= \frac{1}{2}\sum_{i,j=1}^n|\pa_i\mathbf{m}\times\pa_j\mathbf{m}|^2,\quad\mathcal{E}_2 = |\nabla\mathbf{m}|^2,\quad\mathcal{E}_0=1-m_3,
\end{align}
valid for $t_1,t_2$ satisfying the constraints \eqref{constraints}.

\section{Two-dimensional skyrmions}

We now specialise to dimension $n=2$.  In this case solitons are skyrmions, characterised by their degree (or generalised winding number) $N\in\mathbb{Z}$.  If $f(\mathbf{m})$ is any real-valued function on the two sphere then the degree can be calculated using
\begin{equation}
N[\mathbf{m}]\left(\int_{S^2}f\dd A\right) = \int_{\RR^2}f(\mathbf{m})(\mathbf{m}.\pa_1\mathbf{m}\times\pa_2\mathbf{m})\,\dd^2\mathbf{x}.
\end{equation}

In a two-dimensional context the right hand side of inequality \eqref{general bound} is known as the baby Skyrme energy.  A lower bound on this energy in terms of $N$ is presented in \cite{pz95}; we re-derive this argument.

The $\mathcal{E}_2$ term in \eqref{general bound} is straightforwardly bounded from below using \eqref{E4 bound}:
\begin{equation}
\int \mathcal{E}_2\,\dd^2\mathbf{x} \geq 2\int |\mathbf{m}.\pa_1\mathbf{m}\times\pa_2\mathbf{m}|\,\dd^2\mathbf{x} \geq 8\pi N.
\end{equation}
The remaining terms are bounded as follows:
\begin{align}
&\int \Big[ 2(1-t_1^2)\mathcal{E}_4 + (H-t_2^2)\mathcal{E}_0\Big]\,\dd^2\mathbf{x} \nonumber\\
&= \int \Big(\sqrt{2(1-t_1^2)}\sqrt{\mathcal{E}_4} - \sqrt{H-t_2^2}\sqrt{\mathcal{E}_0}\Big)^2\,\dd^2\mathbf{x} \nonumber \\
&\qquad\qquad
+ \int 2\sqrt{2(1-t_1^2)(H-t_2^2)}\sqrt{\mathcal{E}_0\mathcal{E}_4}\, \dd^2\mathbf{x} \\
&\geq 2\sqrt{2(1-t_1^2)(H-t_2^2)} \int (1-m_3)^{\frac12}|\mathbf{m}.\pa_1\mathbf{m}\times\pa_2\mathbf{m}|\,\dd^2\mathbf{x} \\
&\geq  \frac{32\pi}{3} \sqrt{(1-t_1^2)(H-t_2^2)}  N,
\end{align}
where we used that $\int_{S^2}(1-m_3)^{\frac12}\dd^2\mathbf{x} = 8\pi\sqrt{2}/3$.  Combining the above, we obtain that
\begin{equation}
E[\mathbf{m}]\geq 4\pi \left( 2t_1t_2-1+\frac{8}{3}\sqrt{(1-t_1^2)(H-t_2^2)}\right)N[\mathbf{m}].
\end{equation}

This gives a family of bounds, parametrised by $t_1,t_2$ satisfying inequalities \eqref{constraints}.  We now seek to choose $t_1,t_2$ so as to obtain the strongest possible bound.  To this end, let us write $t_1=\sqrt{u/v}$, $t_2=\sqrt{uv}$ for $u,v>0$.  Then the bound can be written
\begin{equation}
E[\mathbf{m}]\geq 4\pi \left( 2u-1+\frac{8}{3}\sqrt{(1-u/v)(H-uv)}\right)N[\mathbf{m}].
\end{equation}
The function $(1-u/v)(H-uv)$ is maximised with respect to variation in $v$ by the choice $v=\sqrt{H}$.  With this choice of $v$ we obtain
\begin{align}
E[\mathbf{m}]&\geq 4\pi \left( 2u-1+\frac{8}{3}\big(\sqrt{H}-u\big)\right)N[\mathbf{m}] \\
&=4\pi \left(\frac{8}{3}\sqrt{H}-1-\frac{2}{3}u\big)\right)N[\mathbf{m}].
\end{align}
We now maximise in $u$ by making $u$ as small as possible.  Since $u=t_1t_2\geq\frac12$, we choose $u=\frac12$ to obtain
\begin{equation}
\label{2D bound}
E[\mathbf{m}] \geq \frac{16\pi}{3}(2\sqrt{H}-1)N[\mathbf{m}],
\end{equation}
obtained from the solution $t_1=(4H)^{-\frac{1}{4}}$, $t_2=(H/4)^{\frac14}$ to the constraints \eqref{constraints}.  Note that, since $t_1t_2=\frac12$, the $\mathcal{E}_2$ term in \eqref{general bound} plays no role in this bound.  With hindsight, it is better to construct a bound using $\mathcal{E}_0$ and $\mathcal{E}_4$ rather than $\mathcal{E}_2$.

In order to test the effectiveness of the bound \eqref{2D bound} we have computed the energies of one-skyrmions.  As in \cite{lh16}, we employ a hedgehog ansatz
\begin{equation}
\mathbf{m}(r,\theta) = (\sin f(r)\cos\theta,\,\sin f(r)\sin\theta,\,\cos f(r)).
\end{equation}
Within this ansatz the energy \eqref{energy} is
\begin{multline}
2\pi \int_0^\infty \Big[ \frac12 \Big( f''+\frac{f'}{r}-\frac{\sin f\cos f}{r^2}\Big)^2 + \frac12\Big((f')^2+\frac{\sin^2f}{r^2}\Big)^2 \\
-\frac12\Big((f')^2+\frac{\sin^2f}{r^2}\Big)^2 + H(1-\cos f) \Big] r\dd r
\end{multline}
and its Euler-Lagrange equation is
\begin{multline}
0 = r(f''''+f''(1-6(f')^2)+H\sin f) + (2f'''+f'(1-2(f')^2)) \\
-r^{-1}(f''(1+2\cos^2f)+\sin f\cos f(1-2(f')^2)) \\
+ r^{-2}f'(1+2\cos^2f) - 3r^{-3}\sin f\cos f.
\end{multline}
This has a power series solution
\begin{equation}
f(r) = \pi + A_1r + A_2r^3 + O(r^4),
\end{equation}
valid for small values of $r$.  It also has an approximate solution,
\begin{equation}
f(r)\approx A_3e^{-\lambda r}\cos(\omega r)+A_4e^{-\lambda r}\sin(\omega r),
\end{equation}
valid for large values of $r$, in which $\lambda,\omega>0$ satisfy $(\lambda+\ii\omega)^2=-\frac12+\ii\sqrt{H-\frac14}$.  We have solved the Euler-Lagrange equation numerically using a shooting method and using these approximate solutions as initial conditions.  As a check on numerical accuracy, we computed the integral $E_4-E_0$ of $\frac12|\triangle\mathbf{m}|^4-H(1-\cos m_3)$, which according to a virial theorem vanishes on minimisers: the largest value of $|E_4-E_0|$ obtained was $3.2\times 10^{-4}$.

\begin{figure}[ht]
\label{fig1}
\begin{center}
\includegraphics{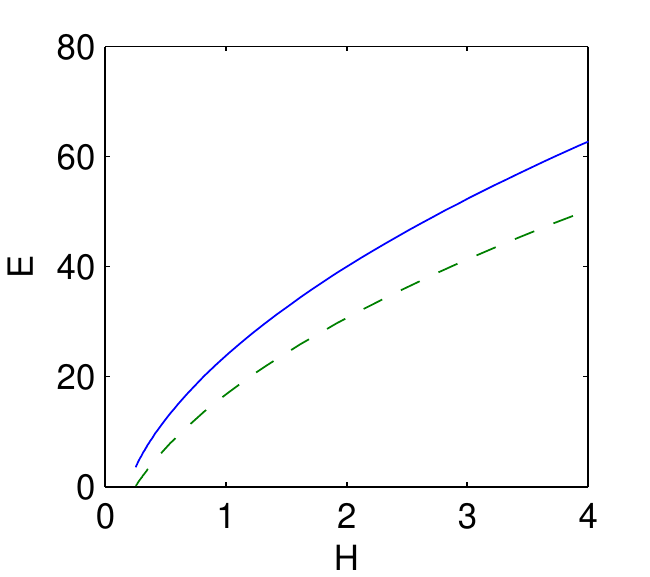}\includegraphics{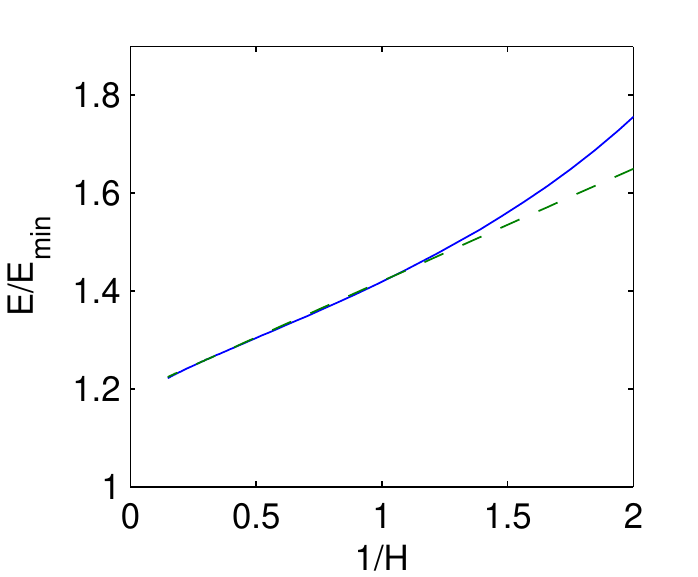}
\end{center}
\caption{Left: energy $E$ (solid curve) of the one-soliton and the lower bound $E_{\text{min}}$ of \eqref{2D bound} as functions of $H$.  Right: the ratio $E/E_{min}$ (solid curve) and the line of best fit $1.18+0.24H^{-1}$ (dashed curve) as a functions of $H^{-1}$.}
\end{figure}

The energies of our solutions are shown in the left plot of figure \ref{fig1}, together with the energy bound \eqref{2D bound}.  One striking feature in figure \ref{fig1} is that, whereas the energy bound approaches zero as $H\to\frac14$, the energy of the skyrmion approaches a non-zero constant $\approx3.4$.  It follows that for small values of $H$ the skyrmion energy is well above the bound.  In contrast, for large values of $H$ skyrmion energies are quite close to the bound.  In the right plot of figure \ref{fig1} we plot the ratio $E/E_{\text{min}}$ of the skyrmion energy to its bound as a function of $1/H$.  The graph is well-approximated by the function $E/E_{\text{min}}\approx 1.18+0.24H^{-1}$, indicating that for large values of $H$ the skyrmion energy is within 20\% of its bound.

\section{Three-dimensional hopfions}

Now we consider the case $n=3$.  In this context the right hand side of \eqref{general bound} is known as the Faddeev energy (or the Faddeev-Skyrme energy), and was introduced in \cite{faddeev75}.  Lower bounds on the Faddeev energy in terms of the Hopf degree $Q$ were obtained in \cite{vk79,harland14,linyang07}.  We will exploit these bounds to obtain bounds on our energy $E$.

Recall that only two of the three terms in \eqref{general bound} played a role in our two-dimensional topological energy bound.  In order to keep the algebra manageable, we restrict our attention here to bounds involving only two of the three terms in \eqref{general bound}.  The Derrick scaling argument \cite{derrick} shows that it is hopeless to try to bound an energy formed from $\mathcal{E}_0$ and $\mathcal{E}_2$, so the two cases to consider are $\mathcal{E}_0$ with $\mathcal{E}_4$ and $\mathcal{E}_2$ with $\mathcal{E}_4$.

We start with $\mathcal{E}_0$ and $\mathcal{E}_4$.  We choose $t_1=\sqrt{s}$ and $t_2=1/2\sqrt{s}$ in order to set the coefficient of $\mathcal{E}_2$ to zero.  The allowed range for $s$ is $1/4H\leq s\leq 1$.  Then our lower bound \eqref{general bound} is
\begin{equation}
E\geq \int \left[ 2(1-s)\mathcal{E}_4 + \left(H-\frac{1}{4s}\right)\mathcal{E}_0\right]\dd^3\mathbf{x}. \\
\end{equation}
In \cite{harland14} it was shown that, for positive constants $\alpha_0,\alpha_4$,
\begin{multline}
\int \left[ \alpha_4\mathcal{E}_4 + \alpha_0\mathcal{E}_0\right]\dd^3\mathbf{x} \\
\geq \frac{8(\alpha_4^3\alpha_0)^{1/4}}{(27\pi)^{1/4}}\left(\int_{S^2}(1-m_3)^{1/6}\,\dd A\right)^{3/2}|Q|^{3/4}.
\end{multline}
Since $\int_{S^2}(1-m_3)^{1/6}\,\dd A=2^{19/6}3\pi/7$, these inequalities together give
\begin{equation}
E \geq \left[(1-s)^3\left(H-\frac{1}{4s}\right)\right]^{1/4}\frac{2^{17/2}3^{3/4}\pi^{5/4}}{7^{3/2}}|Q|^{3/4}.
\end{equation}
This gives a family of lower bounds parametrised by $s$.  The right hand side is maximised by $s = (1+\sqrt{1+12H})/12H$, and this leads to the lower bound
\begin{equation}
\label{hopf bound 1}
E\geq \frac{2^{8}3^{1/2}\pi^{5/4}}{7^{3/2}} \frac{\sqrt{1+12H}-2}{\sqrt{\sqrt{1+12H}-1}}|Q|^{3/4}
\end{equation}

Now we construct a bound using $\mathcal{E}_2$ and $\mathcal{E}_4$.  We choose $t_2=\sqrt{H}$, so that the lower bound \eqref{general bound} is
\begin{equation}
E\geq \int \left[2(1-t_1^2)\mathcal{E}_4 + \left(t_1\sqrt{H}-\frac{1}{2}\right)\mathcal{E}_2\right]\,\dd^3\mathbf{x}.
\end{equation}
The Vakulenko-Kapitanski bound \cite{vk79} (see also \cite{linyang07}) is
\begin{equation}
\int \left[ \alpha_4\mathcal{E}_4 + \alpha_2\mathcal{E}_2\right]\dd^3\mathbf{x}\geq 3^{3/8}16\pi^2\sqrt{\alpha_2\alpha_4}|Q|^{3/4},
\end{equation}
for positive constants $\alpha_2,\alpha_4$.  Combining the above yields
\begin{equation}
E \geq 3^{3/8}16\pi^2\sqrt{(1-t_1^2)(2\sqrt{H}t_1-1)}|Q|^{3/4}.
\end{equation}
The right hand side is maximised when $t_1 = (1+\sqrt{1+12H})/6\sqrt{H}$, and this leads to the lower bound
\begin{equation}
\label{hopf bound 3}
E \geq \frac{2^{9/2}\pi^2}{3^{5/8}} \frac{\sqrt{1+12H}-2}{\sqrt{\sqrt{1+12H}-1}} |Q|^{3/4}
\end{equation}

We have now derived two bounds \eqref{hopf bound 1} and \eqref{hopf bound 3} on the energy of the form $E\geq C(H)|Q|^{3/4}$.  Surprisingly, the coefficients $C(H)$ in the two bounds depend on $H$ in the same way, and differ only by a numerical factor.  The numerical factor in the bound \eqref{hopf bound 1} is $2^83^{1/2}7^{-3/2}\pi^{5/4}\approx 100$ and the numerical factor in the bound \eqref{hopf bound 3} is $2^{9/2}3^{-5/8}\pi^2\approx 112$.  Therefore the second bound \eqref{hopf bound 3} derived using $\mathcal{E}_2$ and $\mathcal{E}_4$ is slightly stronger than that derived using $\mathcal{E}_0$ and $\mathcal{E}_4$.  However, it is conjectured that the bounds on the Fadeev energy from which these are derived can be improved \cite{harland14, ward99}, and improvements in these bounds would improve the bounds \eqref{hopf bound 1} and \eqref{hopf bound 3}.  Of course, the possibility remains that a stronger bound could be obtained using all three terms $\mathcal{E}_0,\,\mathcal{E}_2,\,\mathcal{E}_4$, however, the algebra involved would be tricky.

It is natural to ask whether the power $\frac34$ in the bound \eqref{hopf bound 3} is optimal: does there exist a bound of the form $E\geq C|Q|^\alpha$, with $\alpha>\frac34$?  With minor modifications, a construction due to Lin and Yang \cite{linyang04} shows that there are maps $\mathbf{m}_Q:\RR^3\to S^2$ of Hopf degree $Q$ whose energy is bounded from above by $C'|Q|^{3/4}$ for a constant $C'>0$.  This shows that the power $\frac34$ is optimal.

We first describe the construction of $\mathbf{m}_Q$ in the case $Q=n^2$ for $n\in\NN$.  Fix a continuous function $\mathbf{m}_1:\RR^3\to S^2$ whose first and second derivatives are continuous and bounded, such that $\mathbf{m}_1(\mathbf{x})=(0,0,1)$ outside a compact set.  Fix a continuous function $\mathbf{p}:D^2\to S^2$ on the disc $D^2$ whose first and second derivatives are continuous and bounded, such that $\mathbf{p}=(0,0,1)$ in a neighbourhood of the boundary of the disc, and such that the degree of $\mathbf{p}$ is 1.  Choose $n$ disjoint geodesic discs $B_1,\ldots,B_n$ of radius $\epsilon/\sqrt{n}$ in the lower hemisphere of $S^2$, for fixed $\epsilon>0$ (this can be done as long as $\epsilon$ is sufficiently small).  Construct a map $\mathbf{q}_n:S^2\to S^2$ such that $\mathbf{q}_n$ agrees with $\mathbf{p}$ on each disc $B_i$, and $\mathbf{q}_n=(0,0,1)$ on $S^2\setminus\bigcup_{i=1}^n B_i$.  Extend this function to a neighbourhood of $S^2$ in $\RR^3$ in such a way that $\mathbf{q}_n(\mathbf{y})=\mathbf{q}_n(\mathbf{y}/|\mathbf{y}|)$.  Then it is straightforward to check that
\begin{equation}
\label{derivative estimates}
\sup_{S^2} |\pa_i\mathbf{q}_n|=O(n^{1/2})\text{ and } \sup_{S^2} |\pa_i\pa_j\mathbf{q}_n|=O(n)\text{ for }i,j=1,2,3.
\end{equation}
Unlike in \cite{linyang04}, here we have bounds on second as well as first derivatives.  Now let $\mathbf{m}_Q:\RR^3\to S^2$ be the function
\begin{equation}
\mathbf{m}_Q(\mathbf{x})=\mathbf{q}_n( \mathbf{m}_1(\mathbf{x}/\sqrt{n})).
\end{equation}
This has Hopf degree $Q$ as explained in \cite{linyang04}.

Now we describe the construction of $\mathbf{m}_Q$ in the case $n^2<Q<(n+1)^2$ for suitable $n$.  Let $l=Q-n^2$, and note that $0<l<2n+1$.  Choose $\mathbf{a}_Q\in\RR^3$ sufficiently large such that at every point $\mathbf{x}\in\RR^3$, either $\mathbf{m}_{n^2}(\mathbf{x})=(0,0,1)$ or $\mathbf{m}_l(\mathbf{x}-\mathbf{a}_Q)=(0,0,1)$.  Let
\begin{equation}
\mathbf{m}_Q(\mathbf{x})=\begin{cases} \mathbf{m}_{n^2}(\mathbf{x}) & \mathbf{m}_{n^2}(\mathbf{x})\neq(0,0,1) \\
\mathbf{m}_l(\mathbf{x}-\mathbf{a}_Q) & \mathbf{m}_l(\mathbf{x}-\mathbf{a}_Q)\neq (0,0,1) \\
(0,0,1) & \text{otherwise.}
\end{cases}
\end{equation}
Finally, in the case $Q<0$ we choose $\mathbf{m}_Q(\mathbf{x})=\mathbf{m}_{-Q}(-\mathbf{x})$.


It is straightforward but tedious to check (using the estimates \eqref{derivative estimates}) that there is a constant $C'>0$ such that the energies of the maps $\mathbf{m}_Q$ are bounded above by $C'|Q|^{3/4}$.  Therefore the power $\frac34$ in the bound \eqref{hopf bound 3} is optimal.

\section{Conclusions}

We have obtained topological lower bounds on the energy of two-dimensional skyrmions \eqref{2D bound} and three-dimensional hopfions \eqref{hopf bound 3} in a continuum model of frustrated magnets.  The bounds scale with the degree $N$ and Hopf degree $Q$ as $N$ and $Q^{3/4}$ respectively.  The two-dimensional bound is reasonably strong, in the sense that for large values of $H$ the energy of a one-soliton is roughly 20\% above the bound.  The $\frac34$ power in the three-dimensional bound is optimal, in the sense that the energy does not admit bounds of the form $C|Q|^\alpha$ with $\alpha>\frac34$.  However, it is expected that the coefficient in this bound could be improved substantially by improving bounds on the Faddeev energy \cite{harland14,ward99}.

Although we have presented results for a specific energy functional \eqref{energy}, our methods could be applied to a broader range of energy functionals.  The crucial step in our derivation is that the final line of the inequality \eqref{crucial step} is non-negative for some values of $t_1,t_2$.  So the function $H(1-m_3)$ could be replaced by a function $V(\mathbf{m})$ as long as there exists a unit vector $\mathbf{e}$ and a constant $t_2>\frac12$ such that $V(\mathbf{m})-t_2^2(1-\mathbf{e}.\mathbf{m})\geq0$.  For example, one could consider a potential of the form
\[ V(\mathbf{m}) = H(1-m_3) + \alpha(1-m_3^2) \]
that includes an anisotropy term.  As long as the coefficients satisfy $H>\frac14$ and $H+2\alpha>\frac14$ one could adapt our method to derive a topological energy bound.  On the other hand, a potential with two zeros such as $V(\mathbf{m})=1-m_3^2$ does not seem to be amenable to our method.  Similarly, a potential such as $V(\mathbf{m})=(1-m_3)^2$, whose Taylor expansion $V\approx\frac14(m_1^2+m_2^2)^2$ about its minimum does not include a quadratic term, cannot be tackled using our method.  It would be interesting to know whether either of these potentials supports skyrmions or hopfions.

\medskip
\noindent
\textbf{Acknowledgements} I am grateful to Martin Speight and Paul Sutcliffe for suggesting improvements to a draft.  The results in this paper were conceived during the UK-Japan Winter School in January 2019; I thank the organisers for providing a stimulating environment.

\end{document}